\documentclass{aa}
\usepackage{graphicx}
\def\dg{^{\circ}}
\def\He1211{\mbox{HE\,1211-1707}}

\newcommand{\lppr}{\stackrel{<}{\scriptstyle \sim}}
\newcommand{\lappr}{\raisebox{-0.4ex}{$\lppr$}}
\usepackage{psfig}
\usepackage{epsfig}
\def\lp{LP\,790-29}
\def\gd{GD$\,$229}
\def\grw{Grw+70$^\circ$8247}
\begin{document}

\title{Search for variations in circular-polarization spectra of the 
magnetic white dwarf LP\,790-29\thanks{Based on observations collected 
at the European Southern Observatory, Paranal, Chile (ESO Programme 
65.H-0293)}}

\author{Stefan Jordan\inst1\fnmsep\inst2\fnmsep\inst3
   \and
   Susanne Friedrich\inst1\fnmsep\inst4
}

\offprints{S. Jordan}

\institute{Institut f\"ur Theoretische Physik und Astrophysik, 
             Universit\"at Kiel, 24098 Kiel, Germany
  \and
    Universit\"ats-Sternwarte, Geismarlandstra\ss e 11, 37083 G\"ottingen,
     Germany,
  \and
   Institut f\"ur Astronomie und Astrophysik, Sand 1, 72076 T\"ubingen,
   Germany, jordan@astro.uni-tuebingen.de
  \and
  Astrophysikalisches Institut Potsdam, An der Sternwarte 16,
  14482 Potsdam, Germany,
sfriedrich@aip.de
}

\date{}

\abstract{
We present highly time resolved circular-polarization and flux spectra 
of the magnetic white dwarf \lp\ taken with the VLT UT1  in order to test 
the hypothesis that \lp\ is a fast rotator with a period of the order of 
seconds to 
minutes. Due to low time resolution of former observations this might have 
been overlooked -- leading to the conclusion that \lp\ has a rotational
 period of 
over 100 years. The optical spectrum exhibits one   prominent 
absorption feature with minima at about 4500, 4950, and 5350\,\AA,
which are most likely 
C$_2$ Swan-bands shifted by about 180\,\AA\ in a magnetic field between
50\,MG  and 200\,MG.
At the position of the absorption structures the  degree of  circular
 polarization varies
between -1\%\ and +1\%, whereas it amounts to +8 to +10\%\ in the blue and
red continuum. With this very high degree of polarization
\lp\  is very well suited to  a search for short
time variations, since a variation of several percent in the polarization
can be expected for a magnetic field oblique to the rotational axis.
From our analysis we conclude that variations on time scales from
50 to 2500 seconds must have amplitudes $\lappr 0.7\%$ in the continuum and 
$\lappr 2\%$ in the
strongest absorption feature at 4950\,\AA. 
While no short-term variations could
be found a careful comparison of our polarization data of \lp\ 
with those in the  literatures indicates significant variations on time
scales of decades with a possible period of about 24-28 years. 
\keywords{stars: individual: LP\,790-29 -- stars: rotation 
-- stars: white dwarfs -- stars: magnetic fields }
}
\authorrunning{S. Jordan \& S. Friedrich}
\titlerunning{Search for variations in circular polarization of  LP\,790-29}
\maketitle

\section{Introduction}
\lp\ belongs to a  small group of magnetic white dwarfs which
were suspected to be polarimetrically constant (West \cite{west89};
Schmidt \&\  Norsworthy \cite{schmidt91}) implying 
rotational periods of more than  100 years. 
Alternatively,  the non-detection of variations 
in the polarization can also mean a rotationally symmetric 
magnetic field geometry or a rotational period too short (P$<$10 
minutes, Schmidt \& Norsworthy \cite{schmidt91}) to be 
time-resolved as yet. The shortest  period of 12 minutes 
found in a (magnetic) white dwarf was measured in RE J0317$-$853
(Barstow et al. \cite{barstow95}; Burleigh et al. \cite{burleigh99})
 meaning either that angular momentum is conserved to a higher degree
or that  the star is the product of merging. Segretain et al. (\cite{segretain97})
predicted rotational velocities of about 1000\,km/sec ($P\approx\,40$ sec)
from their models of merging white dwarfs.

If  angular momentum is completely conserved during stellar evolution, 
rotational periods just above the break-up limit of a 
few seconds are even possible in the case of single white dwarfs.
If any of the strongly magnetic white dwarfs really turns out to be
an extremely fast rotator 
it  may  generate a significant anisotropic moment of inertia and 
thereby gravitational radiation measurable with space interferometers
(Heyl \cite{heyl00}).

 Since such a fast rotation  cannot be excluded from previous observations 
we obtained highly  time-resolved flux and circular-polarization 
spectra of \lp,
continuing our search for short rotational periods in white dwarf. 
Friedrich \& Jordan (\cite{friedrich01}) have  started such a search
 with 
a broad-band photometric study of the linear polarization in the famous 
magnetic white dwarf \grw .  So far, no indications for fast rotation could
be found, which  means that the angular momentum has been 
almost completely lost during stellar evolution.
 Recently, long-term  variations in the polarization
of  \gd\ and G$\,$240-7 
 have been observed, from which a rotational 
period of about 100 years can be deduced (Berdyugin \& Piirola 
\cite{berdyugin99}).
 The puzzling slow rotation of most white dwarf stars
is usually explained
by magnetic braking of the stellar core during the process of bipolar 
 outflows that may produce the observed bipolar planetary nebulae 
(Blackman et al. 2001).   Even without a magnetic field  a 
third dredge-up can very efficiently transport angular momentum 
from the core of an AGB star to the envelope so that the final white 
dwarf essentially stops rotating ($v_{\rm rot}=10^{-3}$ km/sec, 
Driebe \&\ Bl\"ocker \cite{bloecker01}).

With its very strong  wavelength dependence
 of the degree of circular polarization 
 (-1\% to +10\%) \lp\ is best suited for  such a study. 
If  \lp\ had a magnetic field not exactly aligned to its
rotational axis, a variation of several percent can be expected. In
the extreme case of a rotational axis perpendicular to both the observer
and the magnetic field axis the polarization in the continuum could
vary between 0\%\ and 20\%\ in order to account for the mean observed value
in the continuum of about 10\%. 
The only shortcoming of observing this particular object is that 
no reliable quantum
mechanical calculations exist as yet for the C$_2$  molecule features
seen in \lp, but simple estimations explain the shifts of the
Swan bands by 500\AA\ with magnetic fields strengths
between 50\,MG (Bues \cite{bues99}) and 200\,MG
(Liebert et al. \cite{liebert78};  Schmidt et al.\cite{schmidt95}); 
therefore  flux and polarization cannot be compared to theoretical
models  in order to measure the  Doppler 
broadening.

\begin{figure}[htbp]
\includegraphics[width=0.5\textwidth]{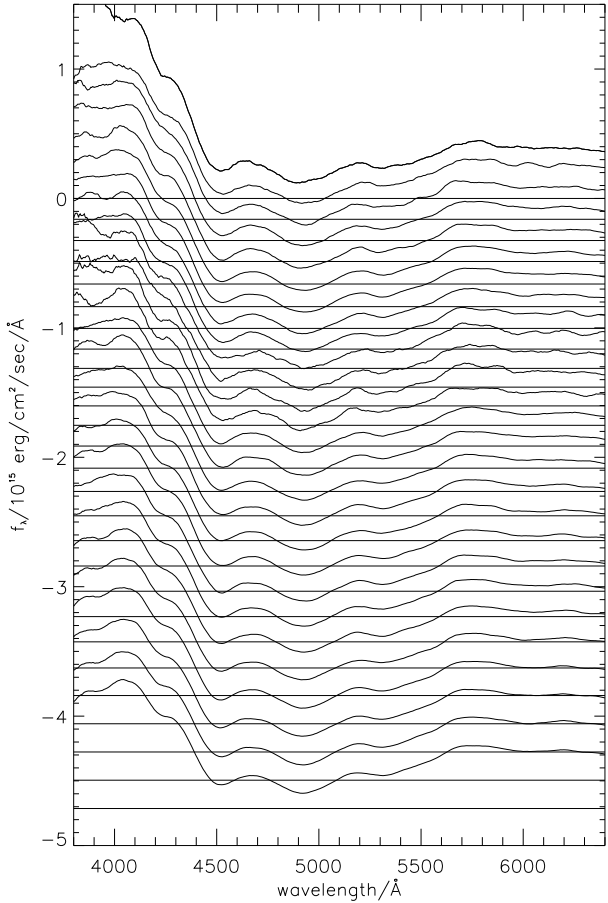}
\caption[]{Phase resolved and Savitzki-Golay filtered flux spectra of \lp\
of the July 3-4, 2000, observing run.
The spectra are shifted proportional to the time past since the first
observation  (total time: 4800 seconds).
The zero point of
the ordinate corresponds to the first (upper) exposure. For all observations
a line is drawn to indicate the zero level.} 
\label{fluxlpall}    
\end{figure}

\begin{figure}[htbp]
\includegraphics[width=0.5\textwidth]{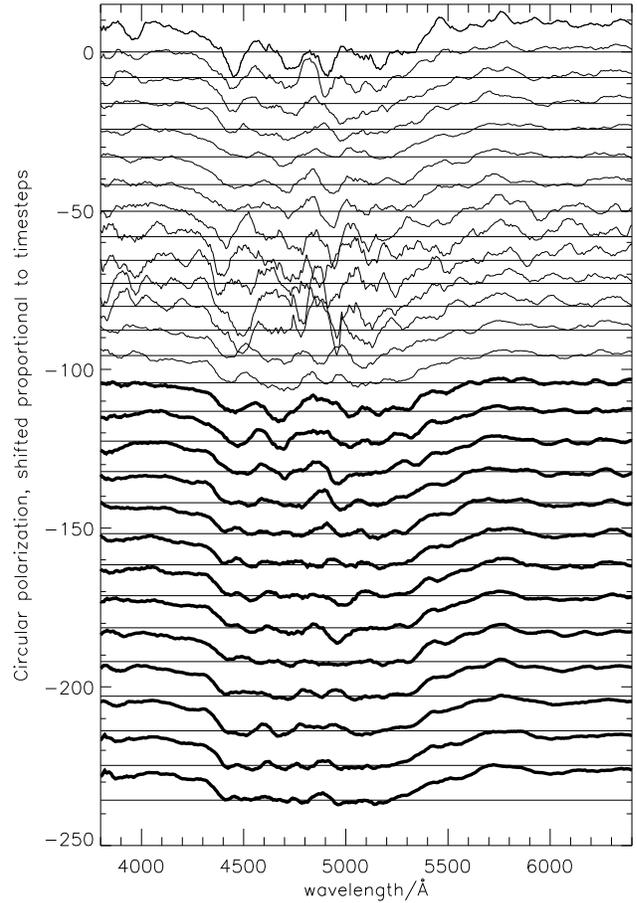}
\caption[]{Circular polarization of \lp\ for 26 phases of the
second observation period covering a total of 4600 seconds.
The noise was reduced with a
Savitzki-Golay filter of  300\,\AA\ width  and a 4th degree polynomial.
The zero points of the degrees of polarization for all observations are 
shown as horizontal lines. Data taken with more than 100 seconds exposure
time are marked with thick lines.
} 
\label{savpollpall}    
\end{figure}

\begin{figure}[htbp]
\includegraphics[width=0.5\textwidth]{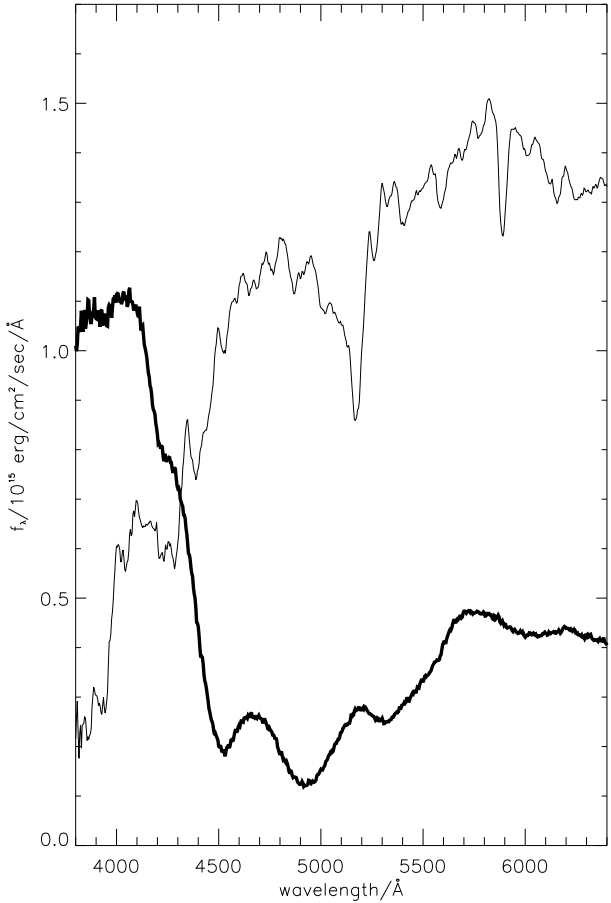}
\caption[]{Mean flux spectra  of \lp\ and of the unpolarized
comparison star.} 
\label{fluxm}    
\end{figure}

\begin{figure}[htbp]
\includegraphics[width=0.5\textwidth]{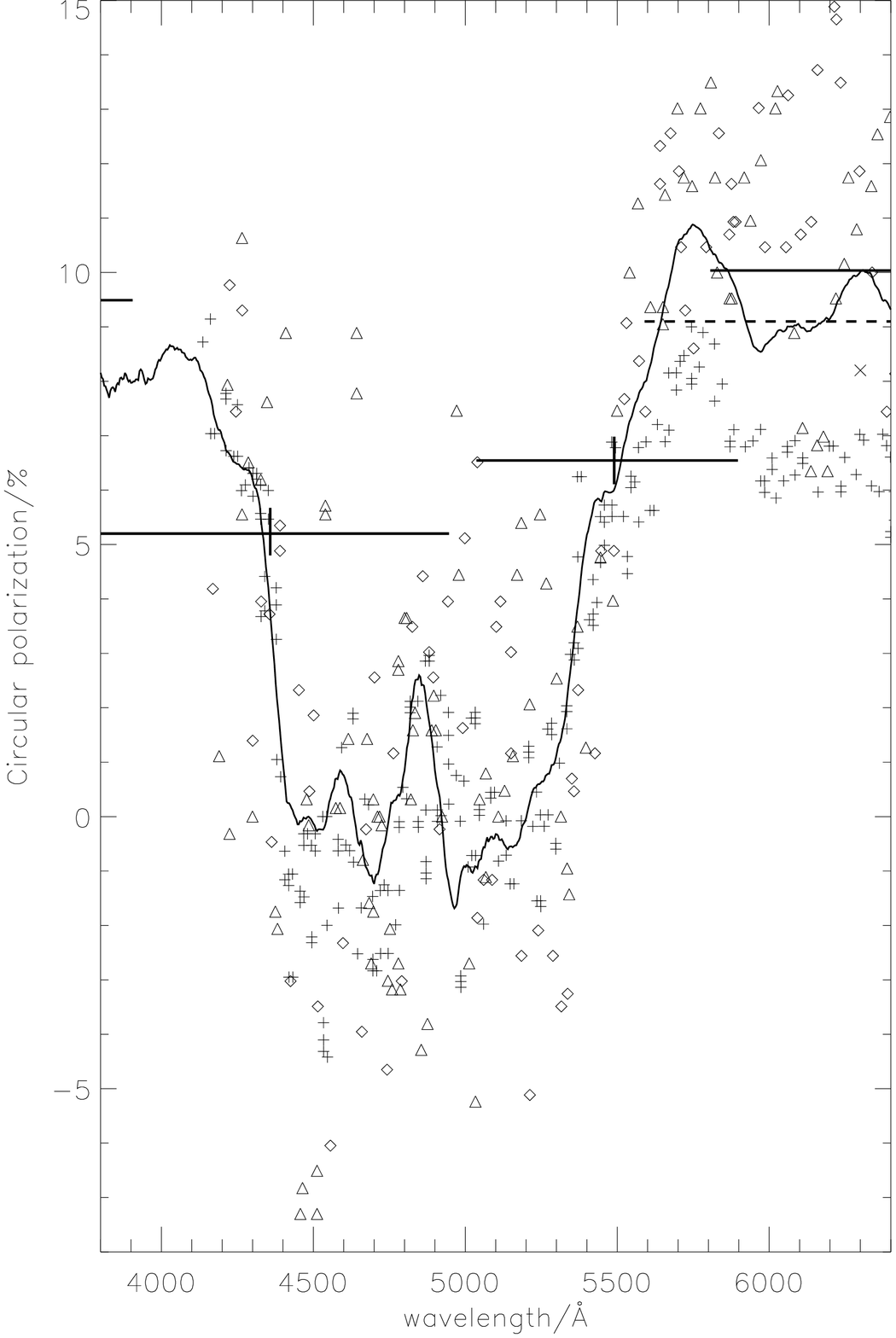}
\caption[]{Comparison of the  mean of our VLT measurements of
circular polarization  (curve) to data from the literature: February 22, 1977
(triangles),  and February 23, 1977 (diamonds), taken by  Liebert et al.
(\cite{liebert78}), May 7$-$8, 1994, (plus signs) observed by  Schmidt et al.
(\cite{schmidt95}). For a clearer distinction these measurements were 
not connected.
The  broad-band data by West (\cite{west89}), marked  by
error bars, were taken between January, 1986, and September 1988,
the measurement in the red by Robert \&\ Moffat (\cite{robertmoffat89},
 February 23, 1987, 'x' symbol)  and  by  Beuermann \&\ Reinsch
(\cite{beuermann01}; dashed line) taken on February 4, 2000.
 } 
\label{secular}    
\end{figure}
  
\section{Observation and data reduction}
Observations of \lp\ were performed on May 5, July 3 and 4, 2000, in service 
mode by ESO staff members with VLT FORS1  in standard resolution. 
Grism GRIS\,150I+17 was used without an order sorting filter to cover the 
spectral range below 6500\AA\ with a resolution of 5.4\AA /pixel. A 
Wollaston prism and a quarter wave plate were inserted to obtain circular 
polarization spectra. The latter was rotated by 90$\dg$
after each consecutive exposure to account for errors in flatfielding, 
offsets from 
the retarder plate zero angle, and wavelength dependencies of the
retardation of the quarter wave plate. 
Flat field, bias, and wavelength calibration exposures were taken with the 
same instrument configuration after each observing night.  Sky spectra 
were taken from regions on the CCD adjacent to the observed stars and 
subtracted from the object spectra.
One anonymous field star at a distance of $1^\prime$ to \lp,
assumed to be  non-variable, was simultaneously observed,
to check for  observationally introduced periods and to obtain
a measure for the accuracy of the polarization data.
Exposure times were  varied between 11s and 83s, in order to 
avoid aliases in periodograms as a consequence of a regular observing scheme. 
In order to reduce read-out time and overhead of the CCD only a window of 
650 pixel $\times$ 2048 pixel was read out. Finally a flux standard (GD108) 
was observed  for a rough flux calibration.

Data reduction was performed with standard IRAF routines. The circular 
polarization of an individual exposure was determined as the difference between
the two  object spectra divided by their 
sum. The final circular polarization  (corrected for observationally 
introduced errors) was determined from two consecutive exposures whose 
retarder angle differ by 90$^\circ$.

During the first observing run on  May 5 \lp\ was observed  9 times
with effective exposure times between 22 and 90 seconds; 
two consecutive exposures with different retarder angles were
used together to calculate the polarization.
Probably due to the atmospheric conditions the measured flux of the comparison
star was highly variable; it  had  a flux level up to a factor  three lower
 than in the second period in July, where the star
 varied by less than 10\%. After 20 minutes the May observing
run was cancelled. Due to the worse exposure 
the photon noise in the polarization  had  strongly increased so that
we based  our  analysis almost entirely
on the data from the July observations. In total we obtained only about
half of the exposure time originally granted to this project,
but if the polarization amplitude  were  a large fraction of
the  10\%\ expected for a 
perpendicular rotation this would be sufficient to measure a period.
 The effective 
exposure times for the 26 observations during the July run vary
between 22 and 166 seconds.
 
\section{Data analysis}
In Fig.\,\ref{fluxlpall} and \ref{savpollpall} we have plotted the
 flux $f_\lambda$ and the circular polarization for all measurements
of the second observing run. 
 Our unsmoothed polarization data (5.4\AA /pixel) have a mean noise
level of about $\pm 6$\% --  too large to directly search
for  variations in the data.
 Therefore, the observations shown in Fig.\,\ref{savpollpall} had
to be smoothed. Since we did not want to broaden potential structures
in the polarization spectrum we used a  Savitzki-Golay filter
(300\,\AA\ width  and a 4th degree polynomial; Press et al. \cite{press92})
which, compared to simple Gaussian or boxcar filters,  
preserves
the height and width of narrow features, still considerably reducing the
noise.

On first glance the polarization data  seem to
indicate large variations which, however, are mostly present
in the observations taken with less than 100 seconds exposure time
(Fig.\,\ref{savpollpall}).
Without a careful inspection it is difficult to judge whether
the variations in the polarization can be trusted or whether these
changes are entirely due to the higher noise level which has
only been partly reduced by the filtering.

For this reason we tried to obtain a more quantitative measure for
the signal-to-noise ratio. We were following two different approaches:
Firstly, variations of the measured degree of circular polarization
in the comparison star should be pure noise. Therefore we
determined the standard deviation $\sigma$ from zero for all observations of
the calibration star. We found that $\sigma \cdot \sqrt{t_{\rm exposure}}$ 
was nearly a constant so that  the noise can approximately
be described by Poisson statistics. We estimated
  $\sigma_{\rm LP}(\lambda)$ in \lp\ by multiplying 
$\sigma_{\rm comp}(\lambda)$  with the square root of the flux ratio
$\sqrt{f_{\lambda,{\rm comp}}/f_{\lambda,{\rm LP}}}$. 

Secondly, we determined $\sigma_{\rm LP}(\lambda)$ without reference to
the comparison star:  For several wavelength intervals we  used the standard
deviation of the  differences between the data and 
the Savitzki-Golay filtered observations  as a measure for the noise.
We concluded that the $\sigma_{\rm LP}(\lambda)$ estimated with the
second method was   about one third larger than with the first
method, probably an indication for small non-Poisson 
contributions.

 We used the larger value for a simulation: We added artificial 
Gaussian noise with the  $\sigma_{\rm LP}(\lambda)$ 
estimated above to the mean polarization spectrum 
and compared the resulting synthetic spectra to the observation.
In both data sets  the variations are of the same order. 
We also could not find any significant differences between the May and 
July observation runs, and therefore we conclude that variations 
 of more than 2\%\ on 
time scales of minutes or months are unlikely.

\section{Period search}
 Nevertheless we tried 
to search for periodicities  using  both the analysis of
variance method by  Schwarzenberg-Czerny (\cite{schwarzenberg89}) and
the phase dispersion minimization  algorithm by Stellingwerf 
(\cite{stellingwerf78}); no significant
differences were found between the results of the two methods.

We searched for periods between 50 and 2500 seconds in 
(a) the mean polarization at the wavelengths of  the strong absorption
structure between
4300 and 5400\,\AA, (b) the continuum from 5800 and 6500\,\AA,
(c) for every wavelength of the Savitzki-Golay smoothed data in
steps of 10\,\AA, and  (d) particularly at 4950\,\AA, where
the deviations are strongest  (about 2\% compared to 0.7\% for 
$\lambda > 5300$\AA).
All attempts to search for periods were applied to both \lp\ and the
comparison star 
in order to eliminate  apparent periods originating from the observing scheme.

 In the periodograms of all four methods peaks occur in the period 
range between 1100 to 1400 seconds. This might be due to the fact that 
the longer-exposed 
 data show a  very approximate repetition   after
1100-1400 seconds. Since no systematic variation consistent with this 
period exists at other wavelengths we conclude that a period 
between 1100 to 1400 seconds is not significant; all variations are
compatible with the assumption that they are due to
residual noise not entirely suppressed by the filtering process.

 Finally, we folded the data with trial periods between 50 and 2500 
seconds (in steps of 10 seconds), repeated the folded spectra 
until they covered the total observing  time 
interval and compared them
to the original polarization data.  
No convincing similarities were found for any period 
so that we conclude that any 
actual short time variations should be smaller than about 
$\lappr 0.7\%$ in the continuum and $\lappr 2\%$ in the
strongest absorption feature at 4950\,\AA .

\section{Search for secular variations}
After having found no clear indications for short term variations
the question is whether secular variations over many years exist.
The hypothesis that \lp\ has a period longer than 100 years stated
by Schmidt \& Norsworthy (\cite{schmidt91}) is entirely based
on two spectra taken in 1977 (Liebert et al. \cite{liebert78}) 
and broad band measurements by West (\cite{west89}) taken between 1986
and 1988.  The measurement by Robert \&\ Moffat (\cite{robertmoffat89}, in
the red part of the optical spectrum) seems to support this hypothesis.
However, a close inspection of the data in the red continuum of \lp\
between  5500 and 6400\,\AA\  shows slight discrepancies:
a circular polarization of 10-13\%\ in the Liebert observations, 
10\% in  West's,
and 8.2\%\ in   Robert's \&\ Moffat's photometry. If we add the
spectro-polarimetric data by Schmidt et al. (\cite{schmidt95}) from
1994, we find that their value for the circular polarization in
the red continuum (6-8\%) clearly differs from the 
Liebert et al. (\cite{liebert78}) data by 4-5\%. 
All data from the literature are plotted in Fig.\,\ref{secular}
together with our VLT  time averaged spectro-polarimetry  
and ESO 3.6m broad-band
polarimetry  by Beuermann \& Reinsch (\cite{beuermann01}). 

The scatter of the polarization  near the strongest
absorption feature (4500-5500\,\AA) is too large to draw any definite
conclusion. However, in the red continuum the star is bright and
 we  conclude  that \lp\ is indeed varying slowly but on a time
scale  smaller than previously stated. In the
wavelength interval between 5500 and 6400\,\AA\ the polarization
changed from $11.5 \pm 1.$5\%\ in 1977 to about 8-10\%\ in
1986/1987, and decreased to $7\pm 1$\%\ in 1994. In 2000,
we  again measured the level of the 1986/1987 polarization;
thus it is possible that about  half a period has been past since then. If
our idea is correct, the rotational period would be about 24-28 years
implying that a higher degree of polarization will be reached again
in 2010-2014. 
 Long term monitoring is necessary to 
clarify the situation.

\section{Discussion}
 Up to now, no changes of
the degree of circular polarization of \lp\ have been reported, resulting in
the conclusion that the rotational period exceeds 100 years. 
As an alternative to such extremely slow rotation we have tested 
highly time-resolved  circular-polarization  spectra
 obtained
with the VLT for variations on time scales of 50 up to 2500 seconds.
In the extreme case of a rotational axis perpendicular to both the 
observer and the magnetic field axis the polarization in the continuum could
vary by about $\pm 10$\%. However, we found  no evidence for  rapid 
variations exceeding 0.7\% in the continuum or 2\%\ at the wavelengths 
of the absorption features. Smaller variations showing up
in our observations are compatible with the assumption that they are
produced by statistical noise. 

The result is also compatible with
fast polarimetric $R$-band photometry (5700-7400\,\AA) by Beuermann \&\
Reinsch (\cite{beuermann01}) who found no significant periodicities 
between 4 seconds and 1.5 hours  in their complementary study of \lp\ which 
extends to longer continuum wavelengths and looks for even shorter 
rotational periods.

 Rapid variations
cannot completely be excluded  on a sub-percent level. However, since
Friedrich \&\ Jordan (\cite{friedrich01}) also could not measure any
fast variations 
in the case of \grw\
fast-rotation as a general scenario for  magnetic white dwarfs
with apparently constant flux and polarization  becomes more and
more improbable.

The quality of  our mean circular-polarization 
spectrum considerably exceeds any previous data of \lp. We compared
it to all other measurements of circular polarization published in the
literature and found that both the first data  of this object
taken in 1977 by  Liebert et al. (\cite{liebert78}) and data
obtained in 1994 by Schmidt et al. (\cite{schmidt95}) were discrepant from
our newer data in the continuum at $\lambda > 5300$\AA.  Together with
broad band data by West (\cite{west89}) from 1986-1988, Robert \&\ Moffat
(\cite{robertmoffat89}) from 1987, and   Beuermann \&\ Reinsch (\cite{beuermann01}) we conclude 
that  the degree of circular polarization actually changes systematically
and that there is a slight indication for a  period between 24 and 28 years.
 Taking this period at face value the polarization at 
$5300-6300$\AA\ should increase for the next 9-13 years. 
Since the question how much of the angular momentum is lost during 
stellar evolution is of  crucial importance it may be worth  
monitoring \lp\ (and other magnetic white dwarfs) with large telescopes 
and long-exposure polarization spectra.  For this purpose one 
high signal-to-noise circular-polarization spectrum (of one or two
hours exposure time) with the  
VLT every year  would be sufficient.

\begin{acknowledgements}
We thank the ESO staff on Cerro Paranal
for observing \lp\ with the VLT in service mode.
Work on magnetic white dwarfs in Kiel is supported by the DFG under 
KO-738/7-1. S. Jordan thanks K. Beuermann for financing his research
in G{\"o}ttingen from his DLR grant 50 OR 9903 6.
\end{acknowledgements}

\end{document}